%%%%%%%%%%%%%%%%%%%%%%%%%%%%%%%%%%%%%%%%%%%%%%%%%%%%%%%%%%%%%%%%%%
%
% VERSION LOG
%
% tp+mnr:	initial version (mid-May)
% chema:	May-23
% tp:		May-25
% tp:		May-31 (Springer version)
% mnr+ms:	Jun-07
% chema+tp:	Jun-08
% mnr+tp:	Jun-12
% chema:	Jun-23 (minor text revisions, H for model A corrected)
%
%%%%%%%%%%%%%%%%%%%%%%%%%%%%%%%%%%%%%%%%%%%%%%%%%%%%%%%%%%%%%%%%%%
%
% DOCUMENT LAYOUT
%
% Lecture Notes in Physics (Proceedings)
% Springer-Verlag, Heidelberg
%
\input lecproc.cmm
%
%%%%%%%%%%%%%%%%%%%%%%%%%%%%%%%%%%%%%%%%%%%%%%%%%%%%%%%%%%%%%%%%%%
%
% CUSTOM DEFINITIONS (CLEARLY INDICATE YOUR MODIFICATIONS, if any)
%
\def		\etal		{et~al.}
\def		\aupun		{\hbox{$^{\underline{\rm a}}$.}}
\def		\Marti		{Mart\'{\i}}
\def		\Ibanez		{Ib\'a{\~n}ez}
\def		\beq		{$$}
\def		\eeq		{$$}

\def    	\Mb		{\hbox{$M_{\rm b}$}}
\def    	\Wb		{\hbox{$W_{\rm b}$}}

\def    	\pc		{\hbox{$p_{\rm c}$}}
\def    	\rhoa		{\hbox{${\rho}_{\rm a}$}}
\def    	\vh		{\hbox{${v}_{\rm h}$}}

\def		\Lj		{\hbox{${L}_{\rm j}$}}

\def     \atx#1			{\remark{{\bf #1}}}		% added/new
\def     \ntx#1			{\remark{{\bf #1}}}		% added/new
\def     \ctx#1			{\remark{{\smallescriptfont #1}}} % comment
\def     \rtx#1			{\remark{{\it #1}}}		% removed
\def     \itx#1			{\remark{{\sans #1}}}		% important

\def	\mrm#1			{\hbox{{\rm #1}}}		% Latex 2.09

\def\where#1{\advftncnt$^\star$\begingroup\petit
\parfillskip=0pt plus 1fil
\def\textindent##1{\hangindent0.5\oldparindent\noindent\hbox
to0.5\oldparindent{##1\hss}\ignorespaces}%
\vfootnote{$^{\star}$}{#1\vskip-9.69pt}\endgroup}
%%%%%%%%%%%%%%%%%%%%%%%%%%%%%%%%%%%%%%%%%%%%%%%%%%%%%%%%%%%%%%%%%%
%
%
%
\contribution
{
Bending relativistic jets in AGNs\where{To appear in Proceedings
of "Relativistic jets in AGNs", Cracow, 27-30 May 1997.}
}
\contributionrunning{Bending relativistic jets}
\author{
T.\ts Plewa@1,
J.\ts M{\aupun}\ts {\Marti}@2,
E.\ts M\"uller@3,\newline
M.\ts R\'o\.zyczka@4@1@5,
M.\ts Sikora@1
}
\authorrunning{Plewa et al.}
\address
{
@1Nicolaus Copernicus Astronomical Center, Warsaw, Poland
@2Departamento de Astronom\'{\i}a y Astrof\'{\i}sica, Universidad
de Valencia, Valencia, Spain
@3Max-Planck-Institut f\"ur Astrophysik, Garching, Germany
@4Warsaw University Observatory, Warsaw, Poland
@5also, Member, Interdisciplinary Center for Mathematical
Modelling, Warsaw, Poland
}
\abstract
{
We present simulations of relativistic jets propagating in a
nonuniform medium.  Specifically, we study the bending of jets
propagating obliquely to the vector of the density-gradient. Our
results are applied to the NGC~4258, where such a medium is assumed to
be provided by the atmosphere of the sub-parsec accretion disk tilted
with respect to the original direction of the jet propagation. As a
result, the jet is bent on a scale comparable to the density
scaleheight of the disk atmosphere. The magnitude of the bending
effect is found to be largest for light jets with low Lorentz
factors. The predicted direction of bending is consistent with the
observations.
}
\titleb{}{Introduction}
Begelman \& Cioffi (1989) and Cioffi \& Blondin (1992) demonstrate
that a light jet propagating in a uniform ambient medium can be
confined and collimated by the cocoon pressure.  However, if the
ambient medium is nonuniform then an asymmetric cocoon may develop,
resulting in pressure differences on opposite sides of the beam. It is
conceivable that this effect may be strong enough to cause a
deflection of the beam on a scale comparable to the scale of ambient
medium nonuniformities. This can explain bending of relativistic jets
on parsec -- kiloparsec scales observed in quasars and radio-galaxies
(see e.g. Appl \etal\ 1996, and references therein), or on sub-parsec
scales (e.g. in NGC~4258; Herrnstein \etal\ 1997).  In the former case
the density-gradient is presumably related to the distribution of
matter in the molecular torus (Baker 1997), in the latter case the
density-gradient can be provided by atmosphere of a warped accretion
disk.  In the present communication we apply our numerical simulations
to the latter case, assuming that the velocity vector of the jet gas
at the inlet is misaligned with the density gradient in the atmosphere
of the disk (Fig.\ts 1).
%
% Figure 1
%
\begfig 6.0 cm
\includegraphics{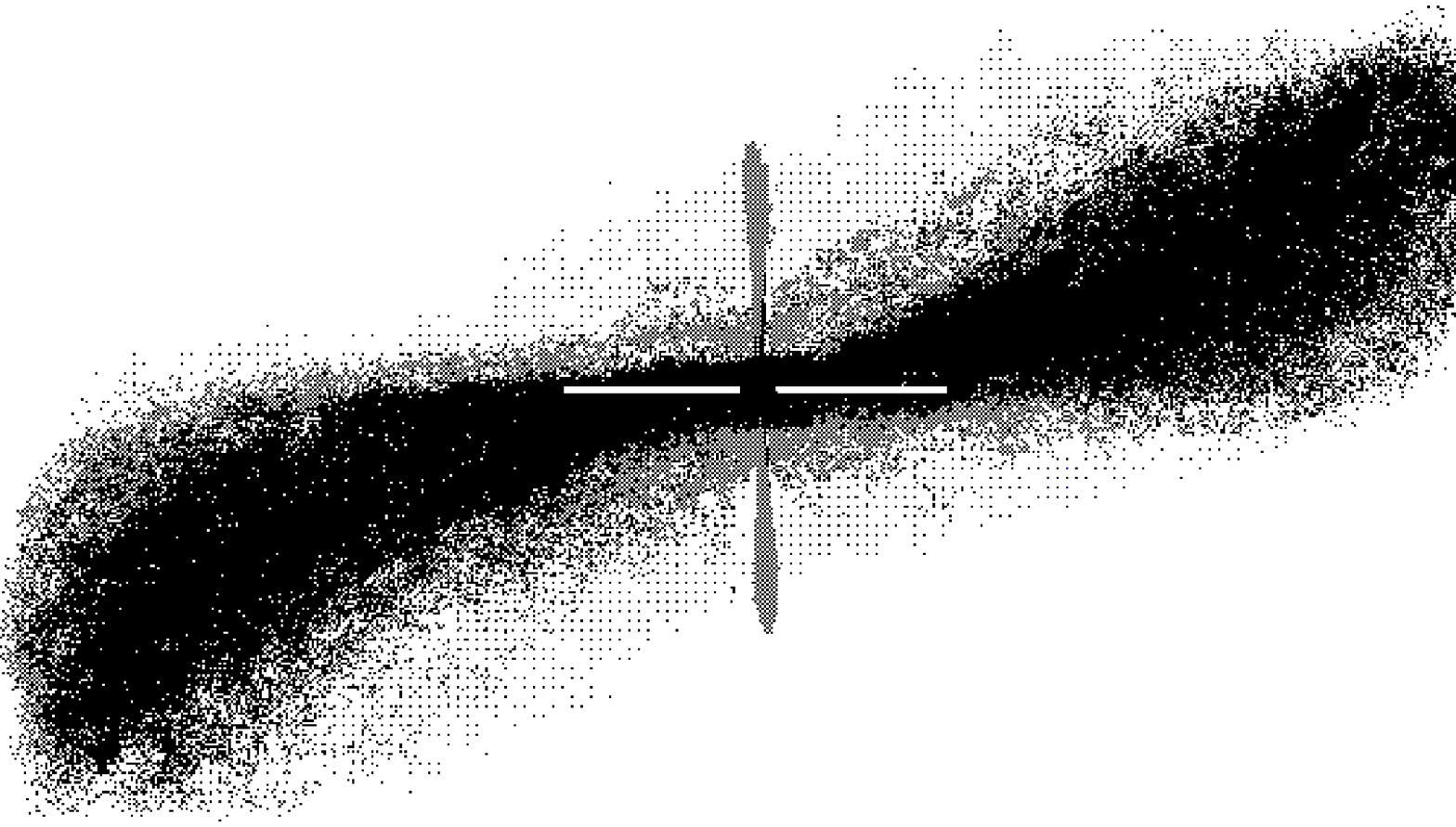}
\figure{1}
{
Schematic representation of initial conditions for jet evolution (see
text for details).
}
\endfig
The observational support for such a scenario is provided by the
warped accretion disk discovered in the central region of NGC~4258
(Herrnstein \etal\ 1997, and references therein). We note that in this
case the perturbations on a sub-parsec scale may be responsible for
the complicated morphology of the jets on larger (up to kpc) scales on
which the jets appear to be composed of several twisted strands
(Cecil, Wilson \& DePree 1985). The strongly warped disk present in
the nucleus of NGC~1068 (Begelman 1997, private communication) may
also influence the radio jet morphology in this object (Gallimore
\etal\ 1996), but the evidence is not yet as compelling as in
NGC~4258.

An estimate of the pressure difference on opposite sides of the beam
may be readily obtained after Begelman \& Cioffi (1989); for a more
rigorous analysis see Loken \etal\ (1992). The cocoon pressure, \pc,
is roughly equal to the total energy deposited by the jet divided by
the volume of the cocoon. Assuming that the energy flux in the jet,
\Lj, and the propagation speed of the jet head, \vh, are constant in
time, and calculating the transversal expansion rate of the cocoon
from the strong shock approximation, for cylindrical jets we obtain
\beq
\pc \propto \left( {\rhoa \Lj \over \vh} \right)^{1/2} t^{-1}, \eqno (1)
\eeq
where \rhoa\ is the density of the ambient medium in which the cocoon
expands. Due to a different dependence of the cocoon volume on time,
the corresponding formula for slab jets reads
\beq
\pc \propto \left( {\Lj \over \vh} \right)^{2/3} {\rhoa}^{1/3} t^{-2/3}. \eqno (2)
\eeq
Thus, an estimate for pressure contrast on opposite sides of the jet is
given by
\beq
{p_c^L \over p_c^R} = \left( {\rho_a^L \over \rho_a^R}\right)^{\alpha},
\eeq
where $L$ and $R$ refer to the left and right side of a slab jet
($\alpha=1/3$), or to points on opposite sides of the symmetry axis of
a cylindrical jet ($\alpha = 1/2$). The pressure contrast can increase
in time as the difference between ${\rhoa}^L$ and ${\rhoa}^R$
increases as the bow shock propagates sideways. However, one may also
expect that the efficiency of the proposed mechanism will decrease
with time due to the general decrease of cocoon pressure.
\titleb{}{Numerical Model}
We perform our simulations with the help of the {\sans AMRA} code
which combines the special relativistic hydro-code of \Marti\ \etal\
(1997) with the AMR (Adaptive Mesh Refinement) approach of
Cid-Fernandes \etal\ (1996). In this study we adopt Cartesian
geometry, and a two-dimensional slab representation for the jets. The
base grid consists of 40 and 60 zones in $x$ and $y$ direction,
respectively, and we use two additional levels of grids with
refinement ratios of 4. Therefore, the resolution of the finest grid
is equivalent to that of a uniform grid of $640\times960$ zones.

The jet material is injected into the computational domain parallel to
the $y$ axis, through the inlet located in the middle of the $x$ axis.
The width of the inlet is equal to 2 zones of the base grid, so that a
resolution of 16 zones per beam radius is achieved on the finest grid.
We use the reflecting boundary condition at the $x$ axis (to the left
and to the right of the inlet), and the transmitting boundary
condition otherwise. The central region of an AGN accretion disk is
approximated by the ambient medium with an exponential density
stratification ${\rho}\sim{\exp(-l/H)},$ where $l$ is the distance
from the mid-plane of the disk. The inlet of the jet is located in the
mid-plane of the disk, and the ambient density is not allowed to fall
below $1\times10^{-4}$ of the mid-plane value. The pressure of the
ambient medium is constant throughout the grid.

In our units speed of light, beam radius, and ambient density at the
inlet are equal to one. The model parameters are the beam Lorentz
factor \Wb, beam Mach number \Mb, proper rest-mass density contrast
(beam to ambient medium) at the inlet $\eta$, pressure contrast $K$,
adiabatic exponent $\gamma$, tilt angle $\theta$ between the mid-plane
of the disk and the $x$ axis, and density scaleheight $H$. Here we
consider only pressured-matched ($K=1$) jets with
$\gamma=5/3$. Further, we restrict our attention to light
($\eta=0.01$), highly--supersonic ($\Mb = 6$) jets which develop large
overpressured cocoons (\Marti\ \etal\ 1997) required for our bending
mechanism to be efficient.

For jets with high internal beam Mach numbers and the same beam
densities, jet luminosity, \Lj, is proportional to
${\Wb}^2$. According to equations (1) and (2), the cocoon pressure is
proportional to \Lj\ raised to the power of 1/2 or 2/3 for cylindrical
and slab jets, respectively.  Therefore, the net bending force (i.e.,
the cocoon pressure gradient) acting on the beam will be proportional
to \Wb\ raised to the power of 1 (cylindrical) or 4/3 (slab). Since
the resistance of the jet to bending is proportional to the momentum
density, which for relativistic flows scales with $W^2$, the bending
mechanism will be less efficient for highly relativistic flows.  Thus,
\Wb\ should not be too large and we adopt $\Wb =6$. Finally, since
tilt angles in excess of 45\degr\ are rather unlikely, we adopt
$\theta=30\degr$.
\titleb{}{Results}
We present three models with different $H$, of which model A ($H=5$,
$\theta=0\degr$) serves as the reference model for models B ($H=2$,
$\theta=30\degr$) and C ($H=5$, $\theta=30\degr$). For these models,
Figs.\ts 2 and 3 present the rest-mass density and gas pressure
distributions at $t=67.5$.

The reference model stays perfectly symmetric throughout the
evolution. Its dynamical structure (left panels of Figs.\ts 2 and 3)
consists of the beam with several internal X-shaped shocks, ending
abruptly at the terminal shock (Mach disk). The beam is surrounded by
the low-density cocoon (light-gray area with an irregular border in
the density plot) which, in turn, is embedded in the ambient medium
swept by the bow shock (the latter is still visible in the upper part
and bottom corners of the computational domain). Because of the
density gradient in the ambient medium, the velocity of the bow shock
and the expansion rate of the cocoon increase with increasing distance
from the mid-plane of the disk, resulting in pear-like shape of the
former and in a roughly conical shape of the latter.

For both models B and C the bending effect and its driving force, the
pressure contrast, are clearly visible. In model B, due to the small
scaleheight, the pressure gradient across the beam is so large that it
causes the beam to decollimate rather than to bend (although the
bending effect is also visible in the upper part of the plot).

A nearly ideal illustration of the expected effect is provided by
model C. The beam is continuously deflected towards the disk axis, but
the internal shocks can be easily identified as counterparts of the
internal shocks in the reference model (right panel in Fig.\ts
3). However, the Mach disk and the shocks above it are irregular,
causing the hottest spot in the head of the jet to wobble about the
beam (the wobbling is best visible in the computer generated movies
accessible at
http://www.camk.edu.pl/$\sim$tomek/RJET/index.html{\#}modelC).
%
% Figure 2
%
\begfig 7.0 cm
\includegraphics{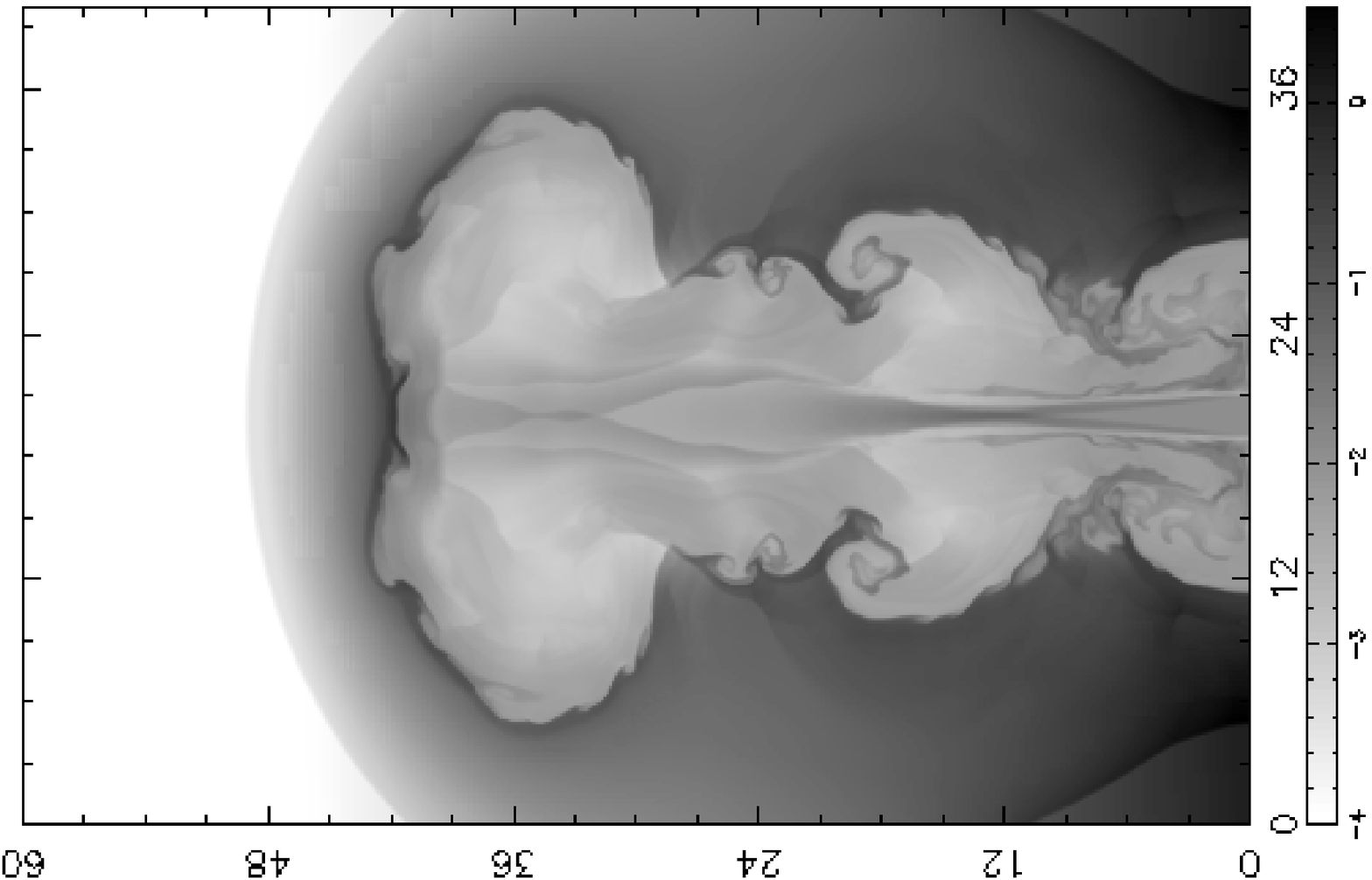}
\includegraphics{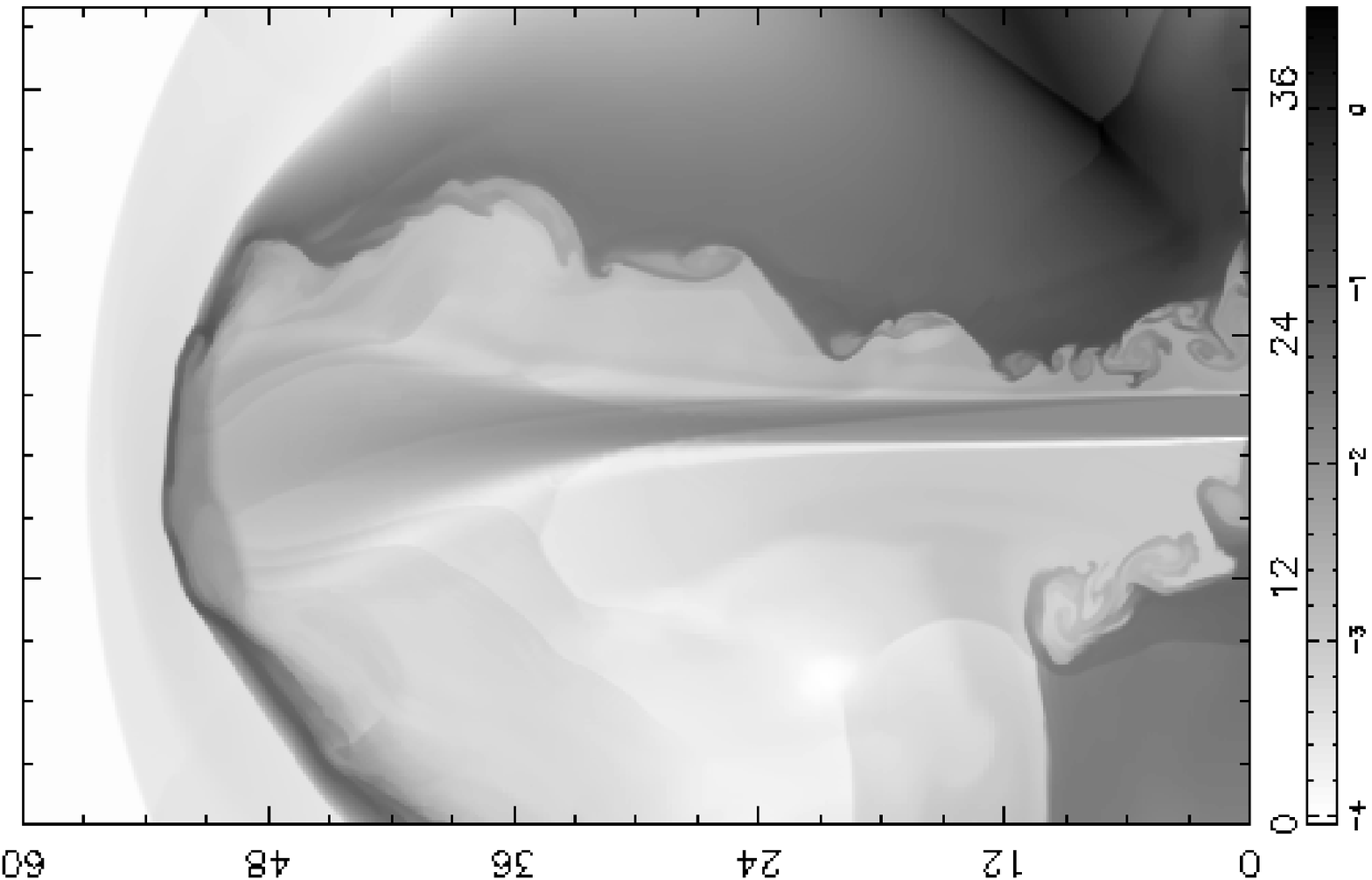}
\includegraphics{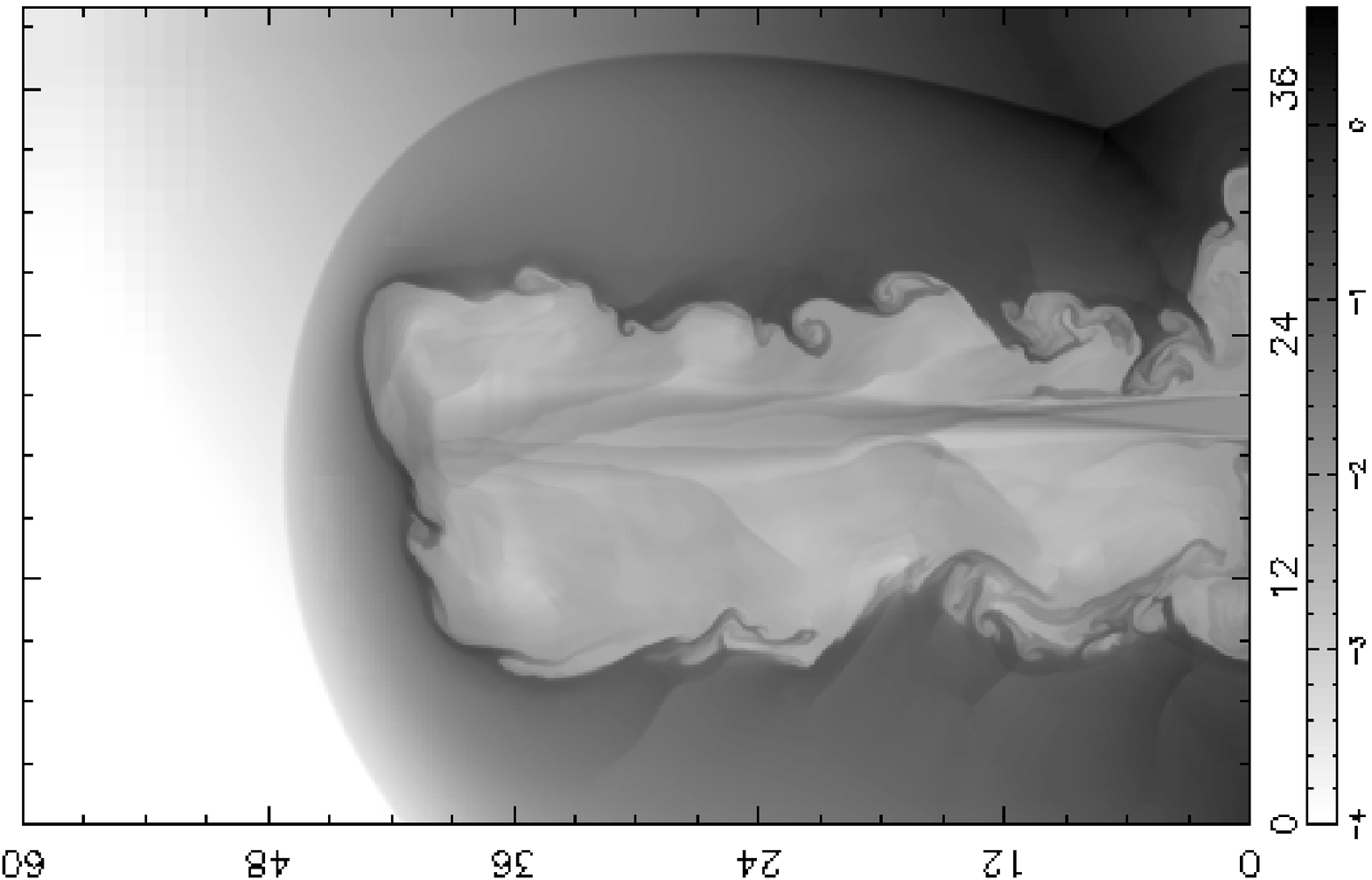}
\figure{2}
{
Density distribution at $t=67.5$.
{\bf left} reference model A ($H=5$, $\theta=0\degr$);
{\bf middle} model B ($H=2$, $\theta=30\degr$);
{\bf right} model C ($H=5$, $\theta=30\degr$).
}
\endfig
%
% Figure 3
%
\begfig 7.0 cm
\includegraphics{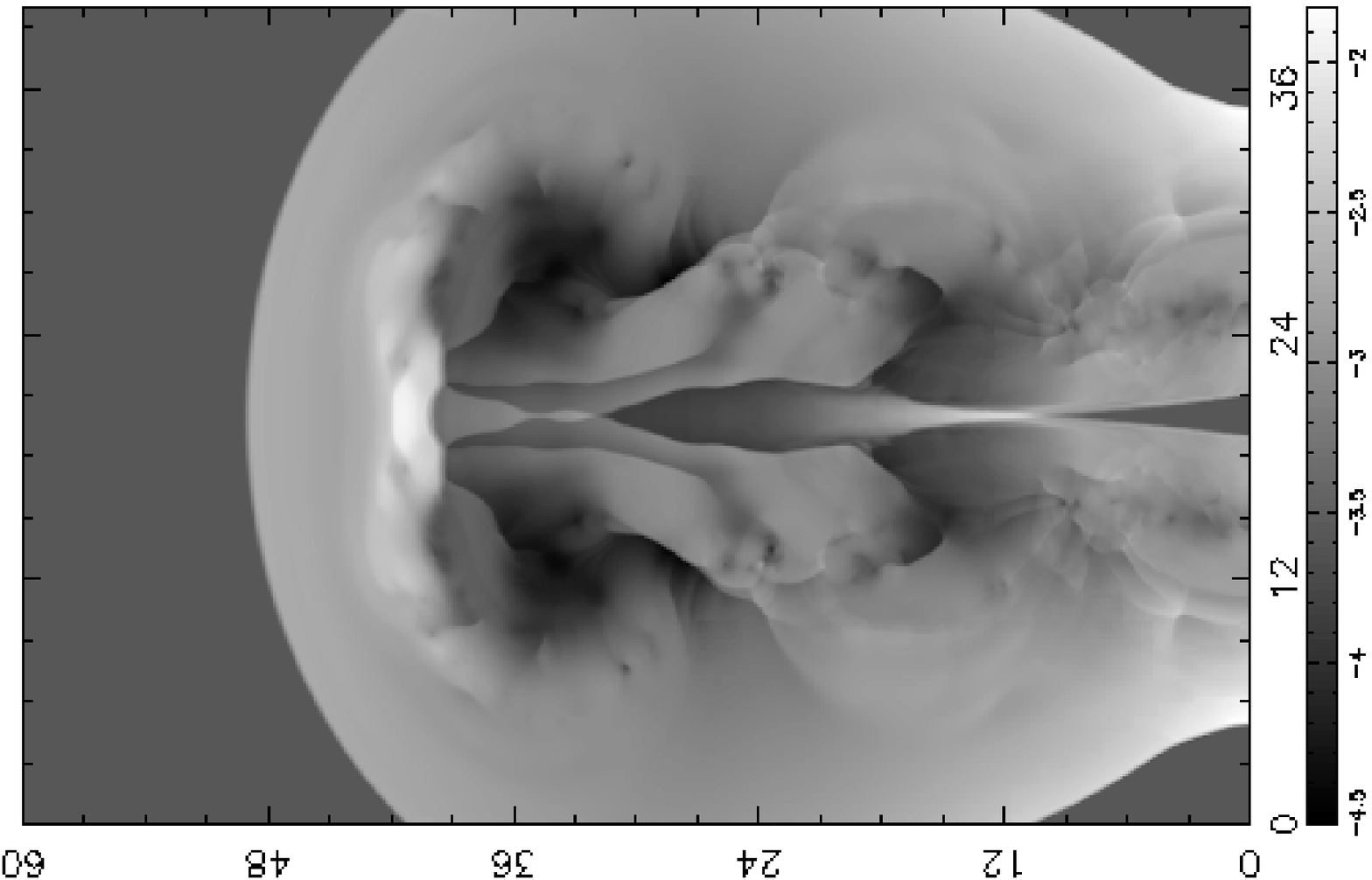}
\includegraphics{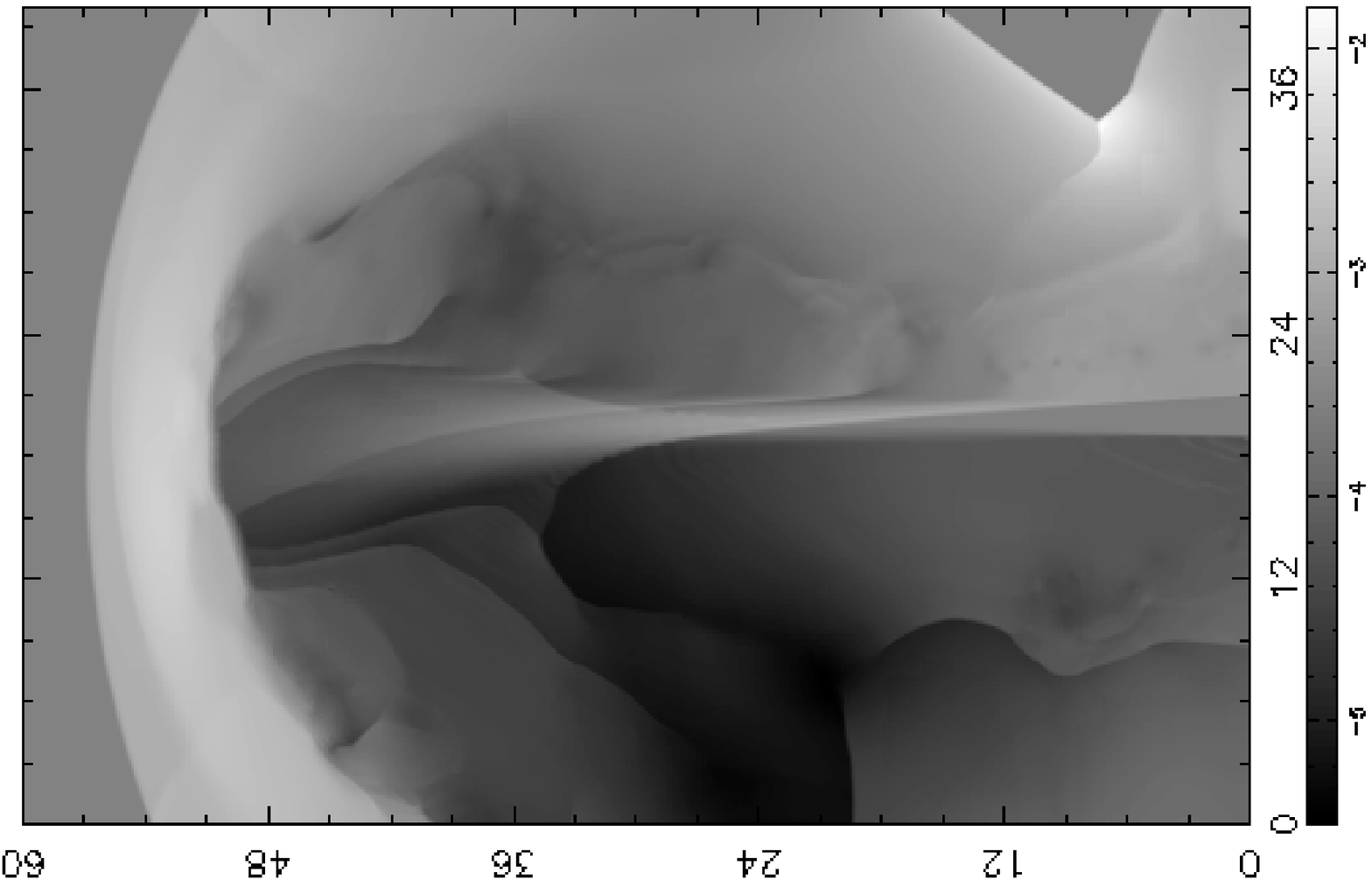}
\includegraphics{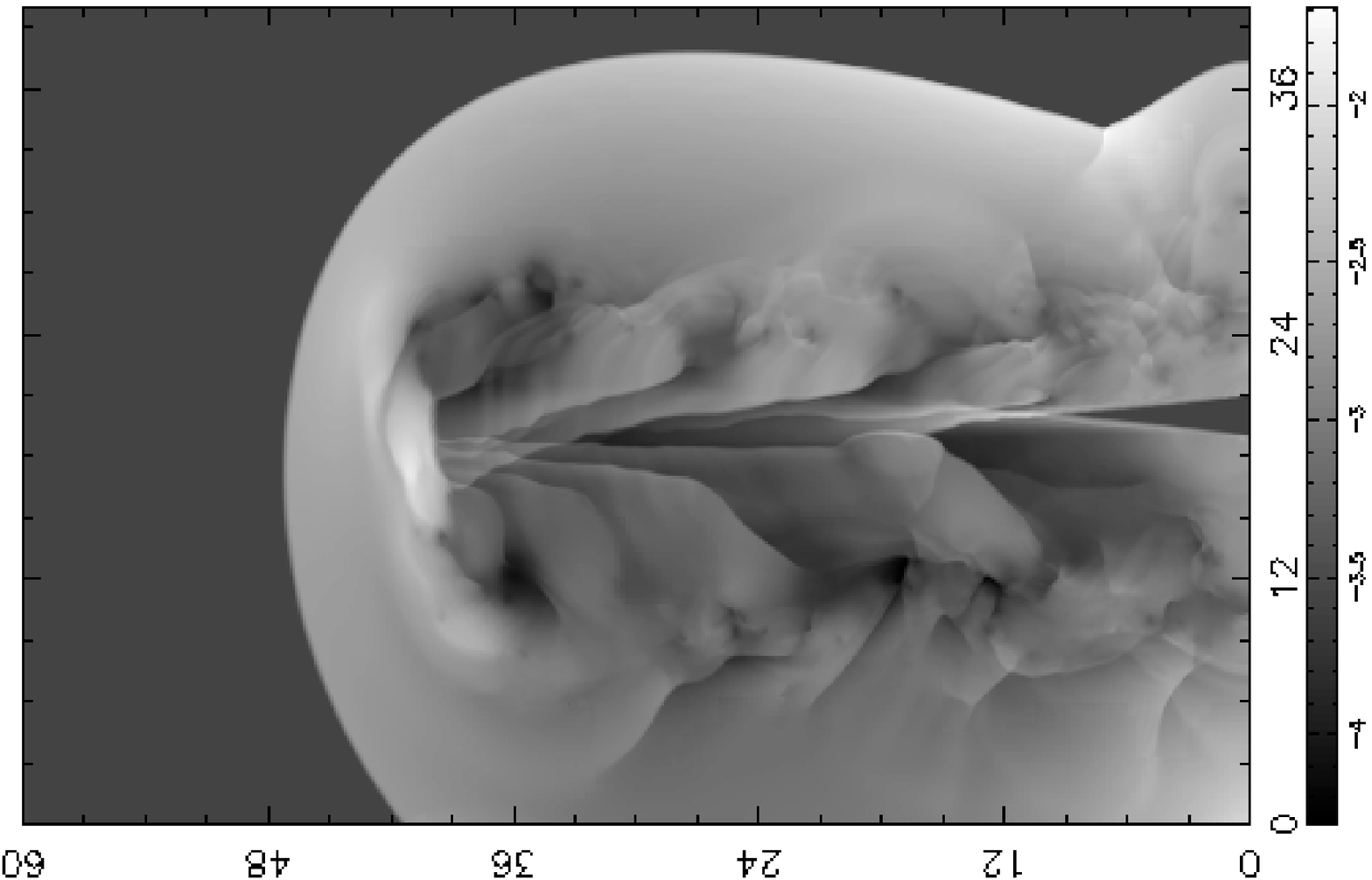}
\figure{3}
{
Pressure distribution at $t=67.5$.
{\bf left} reference model A ($H=5$, $\theta=0\degr$);
{\bf middle} model B ($H=2$, $\theta=30\degr$);
{\bf right} model C ($H=5$, $\theta=30\degr$).
}
\endfig
\titleb{}{Discussion}
Our preliminary results indicate that the proposed bending mechanism
may indeed be efficient enough to cause observable deflections.
According to theoretical estimates, the bending effect will be
particularly strong in the case of light and moderately relativistic
jets. This conclusion is confirmed by a broader survey of the
parameter space (to be reported in near future). Also, we note that
the bending effect can be accompanied by beam decollimation if the
density gradient in the ambient medium is too steep.

Since the time span of our simulations (few months) is rather small,
we cannot be sure if any stationary or quasi-stationary deflections
could be obtained, as the expansion of the bow shock and the cocoon
modifies the density and pressure distributions in the ambient
medium. However, the rotation of the nuclear disk or torus (not
accounted for in our 2D models) would restore the original
distributions. We expect that the restoring effect would be stronger
of the two, but this expectation can only be confirmed by fully 3D
simulations including realistic disk models.

Also, since the cocoon pressure decreases as the cocoon expands, the
net force acting on the jet tends to decrease with time. On the other
hand, the ambient density contrast increases as the bow shock
propagates sideways, tending to strengthen the bending
mechanism. Another comment concerns the flows in the cocoon.  In the
slab jet simulations presented here, the two sides of the cocoon are
disconnected, and pressure balance in the cocoon can only be reached
by pushing the beam sideways. However, in the case of an initially
cylindrical, three-dimensional jet, the flow of gas in the azimuthal
direction (in the cocoon, around the beam), would tend to suppress the
bending mechanism. All these effects should be included in future
investigations.
\titleb{}{Acknowledgements}
The work of TP and MR was supported by the grant KBN 2P-304-017-07
from the Polish Committee for Scientific Research. TP acknowledges the
support from the German-Spanish program ``Accion Integrada''. MS and
TP were supported by the KBN grant 2P-03D-012-09.  JMM has benefited
from a European Union contract (No.~ERBFMBICT950379).  The major part
of the simulations was performed at the Interdisciplinary Centre for
Computational Modelling in Warsaw.
\begrefchapter{References}
\ref Appl, S., Sol, H. \& Vicente, L. 1996, A\&A, 310, 419
\ref Baker, J. C. 1997, MNRAS, 286, 23
\ref Begelman, M. C. \& Cioffi, D. F., 1989, ApJ, 345, L21
\ref Cecil, G., Wilson, A. S. \& DePree, C., 1995, ApJ, 440, 181
\ref Cid-Fernandes, R., Plewa, T., R\'o\.zyczka, M., Franco, J.,
Terlevich, R., Tenorio-Tagle, G. \& Miller, W., MNRAS, 283, 419
\ref Cioffi, D. \& Blondin, J. M., 1992, ApJ, 392, 458
\ref Gallimore, J. F., Baum, S. A., O'Dea, C. P. \& Pedlar, A., 1996,
ApJ, 458, 136
\ref Herrnstein, J. R., Moran, J. M., Greenhill, L. J., Diamond,
P. J., Miyoshi, M., Nakai, N. \& Inoue, M., 1997, ApJ, 475, L17
\ref Loken, C., Burns, J. O., Clarke, D. A. \& Norman, M. L., 1992, ApJ,
392, 54
\ref {\Marti}, J. {M\aupun}, M\"uller, E., Font, J. A.,
{\Ibanez}, J. M\aupun\ \& Marquina, A., ApJ, 479, 151
\endref
\byebye